# TRACING 3-D MAGNETIC FIELD STRUCTURE USING DUST POLARIZATION AND THE ZEEMAN EFFECT

Brandon Shane
Department of Physics, Engineering Physics, and Astronomy, 64 Bader Lane, Queen's University, Kingston, Ontario, Canada, K7L 3N6
and
Department of Physics and Astronomy, Rutgers University, 136 Frelinghuysen Rd, Piscataway, NJ 08854, USA

Blakesley Burkhart
Department of Physics and Astronomy, Rutgers University, 136 Frelinghuysen Rd, Piscataway, NJ 08854, USA and
Center for Computational Astrophysics, Flatiron Institute, 162 Fifth Avenue, New York, NY 10010, USA

Laura Fissel
Department of Physics, Engineering Physics, and Astronomy, Queen's University, 64 Bader Lane, Kingston, Ontario, Canada, K7L 3N6

Susan E. Clark
Department of Physics, Stanford University, Stanford, California 94305, USA and
Kavli Institute for Particle Astrophysics & Cosmology (KIPAC), Stanford University, Stanford, CA 94305, USA

Philip Mocz
Center for Computational Astrophysics, Flatiron Institute, 162 5th Ave, New York, NY 10010, USA

Michael M. Foley
Center for Astrophysics | Harvard & Smithsonian, 60 Garden Street, Cambridge, MA 02138 and
Department of Earth and Planetary Sciences, Harvard University, 20 Oxford Street, Cambridge, MA 02140

## ABSTRACT

The characterization of magnetic fields within molecular clouds is fundamental to understanding star formation processes. Accurately gauging the three-dimensional structure of these fields presents a challenge, as observational techniques such as dust polarization and the Zeeman effect each provide only partial information on the orientation and line-of-sight strength, respectively. By analyzing a suite of AREPO simulations, this paper investigates how observables can relate to underlying physical properties to derive a more comprehensive picture of the magnetic field's inclination angle and strength, specifically in regions where both dust polarization and Zeeman data are available. To demonstrate the method, we produce synthetic observations of the polarization angle dispersion and line-of-sight Alfvén Mach Number and explore the behavior of the inclination angle, $\gamma$, and strength of the magnetic field in regions where both Zeeman and dust polarization data are available. We find that dust polarization data can be used to determine the inclination angle if the cloud is known to be trans-Alfvénic or sub-Alfvénic. The strength of the magnetic field relative to turbulence can be estimated by comparing polarization observations to Zeeman observations. Comparing the dispersion of the polarization angle to the estimated line-of-sight Alfvén Mach Number provides clues about the strength of the magnetic field and, consequently, the orientation of the magnetic field.

## 1. INTRODUCTION

The structure of molecular clouds within the interstellar medium (ISM), and the properties of stars that form within them, are greatly influenced by the presence of magnetic fields (Mestel & Spitzer 1956; Mouschovias 1991; McKee & Ostriker 2007). Magnetic fields affect the dynamics of gas over scales ranging from spiral arms (kpc) to the collapse of individual ($\ll 1\,\mathrm{pc}$) dense molecular cores (Girichidis et al. 2020; Burkhart 2021; Pattle et al. 2023). In particular, supersonic magnetohydrodynamic (MHD) turbulence generates over-densities that can become unstable to gravitational collapse and create protostellar systems (Krumholz & McKee 2005; Federrath & Klessen 2013; Gómez et al. 2018; Shu et al. 1987; Burkhart 2018; Saydjari et al. 2021). Therefore, it is necessary to measure the strength and direction of the magnetic field structure in order to understand the energy balance of the ISM and how MHD turbulence regulates the lifecycle of molecular clouds.

Numerical simulations of MHD turbulence in the context of star formation have provided ample opportunity to study the dynamics and evolution of magnetic fields and how magnetic fields affect star formation (Burkhart et al. 2014; Krumholz 2014; Mocz & Burkhart 2018; Chen et al. 2019). However, observations of magnetic fields are restricted to certain tracers and techniques, all of which provide only partial information on the magnetic field in the molecular and atomic ISM (Crutcher 2012; Lazarian et al. 2015; Haverkorn 2015; Heiles & Haverkorn 2012). One commonly used observational diagnostic of magnetic fields in molecular clouds is the Zeeman effect, which can be used to directly measure the line-of-sight (LOS) component of the magnetic field strength (Crutcher 2012). Other common tracers include the polarization of thermal emission from dust grains and the absorption of background starlight by foreground dust, which can be used to map the plane-of-sky (POS) component orientation of the magnetic field (Davis & Greenstein 1951a; Hildebrand et al. 2000).

Non-spherical dust grains tend to align with their long axes perpendicular to their local magnetic field, which is consistent with the predictions of the theory of radiative alignment torques (RAT, Lazarian 2007; Hoang & Lazarian 2014; Andersson et al. 2015). Starlight that has passed through dust clouds is therefore observed to be polarized in the direction parallel to the magnetic field due to selective extinction (Davis & Greenstein 1951a; Hall 1949) while thermal dust emission is linearly polarized in the direction perpendicular to the magnetic field (Hildebrand et al. 2000). Polarimetric observations of thermal emission from dust at mm/far-IR wavelengths have been used to trace the projected magnetic field orientation on the POS in many Galactic environments (e.g., Hildebrand et al. 1984; Novak et al. 1997; Matthews et al. 2001; Planck Collaboration et al. 2020), and to study the physics of interstellar dust (e.g., Andersson & Potter 2007; Whittet et al. 2008; Cashman & Clemens 2014; Ashton et al. 2018; Hensley & Draine 2023).

Dust polarization gives information on the POS orientation of the magnetic field but does not directly measure the strength of the field. Statistical methods have been introduced in order to indirectly estimate the magnetic field strength from dust polarization observations. These include the Davis-Chandrasekhar-Fermi method and variations on it (Davis & Greenstein 1951b; Chandrasekhar & Fermi 1953; Falceta-Gonçalves et al. 2008; Heitsch et al. 2001; Hildebrand et al. 2009; Houde et al. 2009; Chen et al. 2022). Other techniques, such as the histogram of relative orientation (HRO) technique (Soler et al. 2013; Soler & Hennebelle 2017; Heyer et al. 2020; Barreto-Mota et al. 2021), use the shape of the distribution of relative orientations between the gas structure and polarization vectors to probe the relative importance of the magnetic field, and the polarization intensity gradient method (Koch et al. 2012; Tang et al. 2019).

The most direct method of measuring the strength of the magnetic field in neutral and molecular media is Zeeman splitting. In the presence of a magnetic field, spectral lines will be split into multiple components with the frequency separation between the components proportional to the magnetic field strength (Crutcher et al. 2010; Crutcher 2012). In the interstellar medium, Zeeman splitting can only be used to measure the LOS component of the magnetic field strength. The Zeeman effect has been used to study magnetic fields in the atomic ISM using the H I 21-cm line (Heiles & Troland 2005) and in molecular gas, in particular, using OH (Troland & Crutcher 2008) and CN lines (Falgarone et al. 2008).

There has not yet been a large-scale, detailed statistical comparison between dust polarization and Zeeman measurements. There are a number of complications with using observational data to make this comparison. Dust emission is almost always optically thin and so the polarization measurement would be averaged over the entire column within the LOS probed by the telescope beam. Zeeman measurements are made for individual LOS velocity components and will only sample regions along the column that are emitting a particular spectral line. There are also far fewer Zeeman measurements in general than dust polarization measurements. Additionally, dust polarization surveys and Zeeman measurements may have different resolutions. Many Zeeman measurements are of interstellar absorption lines seen against background radio continuum sources and so give "pencil-beam" measurements of the LOS (e.g., Heiles & Troland 2004; Thompson et al. 2019). Finally, many Zeeman measurements result in non-detections, or detections with a significance $< 3\sigma$ (Crutcher et al. 2010).

Despite these challenges, the LOS sensitivity to the magnetic field provided by the Zeeman detection and POS information from dust polarization could be used to better constrain the overall 3-D magnetic field structure of molecular clouds. By combining data from Crutcher (2012), which collects Zeeman measurements from several previous studies and polarization surveys such as the all-sky polarization maps from the Planck survey at 353 GHz (Planck Collaboration et al. 2015a), James Clerk Maxwell Telescope (JCMT) Pol 2 (Friberg et al. 2016), and the Stratospheric Observatory for Infrared Astronomy (SOFIA) (Harper et al. 2018a) one could ascertain the 3-D magnetic field structure. Additionally, other innovative work has been done to ascertain the magnetic field's 3-D morphology. Tahani et al. (2018) have used Faraday rotation measurements sensitive to the POS magnetic field, Zeeman splitting measurements, and Planck dust polarization observations to model the 3-D magnetic field structure in the envelopes of filamentary molecular clouds. These works are promising and point to the power of join analyses of different tracers.

To properly model the magnetic morphology and its influence on a cloud's structure, the magnetic field's inclination would need to be ascertained. The inclination angle is the angle that the mean magnetic field makes with respect to the plane-of-sky. Estimating the inclination angle of the magnetic field is important as the direction of the magnetic field will influence the physical structure of the cloud along the line-of-sight and will determine what tracers are and are not effective at estimating the magnetic field strength. However, it is not enough to know simply the strength of the magnetic field in absolute terms; its relative importance to the other physics involved can strongly affect cloud dynamics and structure. The Alfvén Mach Number is a metric to determine the relative importance of the magnetic field strength to turbulent motions within a given cloud. The Alfvén Mach Number, $\mathcal{M}_A$, can be written as

$$\mathcal{M}_A \equiv \frac{v}{v_A} = \sqrt{\frac{E_{turb}}{E_B}}, \quad (1)$$

where $E_{turb}$ is the turbulent energy, $E_B$ is the magnetic energy, $v$ is the characteristic velocity and

$$v_A = \frac{B}{\sqrt{4\pi\rho}} \quad (2)$$

is the Alfvén velocity.

In this paper, we compare synthetic observations of dust polarization and Zeeman splitting from AREPO MHD simulations (Mocz et al. 2017) in order to search for methods of constraining the total magnetic field strength, Alfvén Mach Number, and inclination angle of the magnetic field. Our approach builds on previous statistical studies of synthetic dust polarization by Chen et al. (2019), King et al. (2019) and Sullivan et al. (2021) by additionally including synthetic Zeeman splitting observations. In Section 2 we will discuss the simulations used for this analysis. Our implementations of dust polarization and the Zeeman effect are discussed in Sections 3 and 4 respectively. We summarize our method in Section 5 and discuss future work and connections to observations in Section 6. Finally, we summarize this method and our findings in Section 7.

## 2. SIMULATIONS

The simulations used in this analysis were first introduced in Mocz et al. (2017) using the MHD capabilities of the quasi-Langrangian, moving mesh AREPO code developed by Springel (2010) and are available on the Catalogue for Astrophysical Turbulence Simulations (CATS) (Burkhart et al. 2020). They are a suite of solenoidally driven, supersonic, isothermal, magnetized, turbulent boxes with relevant parameters to the molecular clouds on parsec scales. Each simulation solves the ideal MHD equations with an unstructured vector potential constrained transport solver (Mocz et al. 2016). Turbulence is driven in a divergence-free, solendial manner in Fourier space on the largest spatial scales with an Ornstein-Uhlenbeck process (Federrath et al. 2010; Federrath 2015; Bauer & Springel 2012). The simulations have a physical size of 5.2 pc. Each simulation has a sonic Mach number $\mathcal{M}_s \approx 10$ and sound speed $c_s = 0.2$ km s$^{-1}$. The mean magnetic field is characterized by the corresponding average Alfvén Mach Number $\mathcal{M}_{A,0}$ which is the mean initial ratio of turbulent kinetic energy to magnetic energy in the cloud. The average initial magnetic strength in each run is 1.2, 12, 36, and 120 $\mu$G. The mass of each run is 4860 $M_\odot$. The four initial mean Alfvén Mach Numbers of these simulations are listed in Table 1. These simulations span from very weak seed fields (run 1) to strong fields whose magnetic energy density is much greater than the turbulent kinetic energy density (run 4).

For this analysis, we use snapshots taken after self-gravity has been activated following turbulent equilibrium and the first cores collapse in order to accurately mimic self-gravitating molecular clouds in our synthetic observations.

Figure 1 shows column density (grey color bar scale) overlaid with magnetic field lines integrated along the observer's line-of-sight (red lines) for two simulations. The top row shows the line-of-sight where the mean magnetic field direction is entirely in the plane-of-sky. The bottom row shows the line-of-sight where the mean magnetic field direction is parallel to the observer's line-of-sight. The first column shows run 1 of our snapshots which corresponds to a high Alfvén Mach Number. The second column corresponds to run 4 which has the lowest Alfvén Mach Number and the highest magnetic field strength.

**Note.** — [a]Here $B_0$ refers the average initial magnetic field strength in the simulation.

| Run | $\mathcal{M}_{A,0}$ | $\mathcal{M}_s$ | $B_0$ [$\mu$G] [a] | $t_{collapse}$ [$t_{ff}$] | comment |
|---|---|---|---|---|---|
| 1 | 35 | 10 | 1.2 | 0.12 | very weak field |
| 2 | 3.5 | 10 | 12 | 0.16 | weak field |
| 3 | 1.2 | 10 | 36 | 0.17 | moderate field |
| 4 | 0.35 | 10 | 120 | 0.37 | strong field |

**Table 1**
Adapted from Mocz et al. (2017). $\mathcal{M}_{A,0}$ is the initial ratio of turbulent kinetic energy to magnetic energy in the cloud, i.e., the Alfvén Mach Number. $\mathcal{M}_s$ is the sonic Mach number within the cloud. The snapshots of the simulations used for this analysis were taken after the first cores collapsed at $t_{collapse}$ given in units of freefall time $t_{ff}$.

### 2.1. Interpolation

In order to transform our simulations into synthetic Zeeman and dust polarization maps we perform an interpolation of the positional and particle information in the snapshots onto a uniform grid. We performed Smoothed Particle Hydrodynamics (SPH) interpolation. This interpolation closely follows the original SPH algorithm of Springel (2005). First, a Cartesian mesh is initialized at the chosen resolution. This mesh will serve as the base onto which quantities from the moving mesh are interpolated. Using the smoothing length of each moving mesh cell, we calculate how much of a chosen quantity should be deposited into each Cartesian cell using the following weighting prescription:

$$W(r,h) = \frac{8}{\pi h^3} \begin{cases} 1 - 6\left(\frac{r}{h}\right)^2 + 6\left(\frac{r}{h}\right)^3 & : 0 \leq \frac{r}{h} \leq \frac{1}{2} \\ 2\left(1 - \frac{r}{h}\right)^3 & : \frac{1}{2} < \frac{r}{h} \leq 1 \\ 0 & : \frac{r}{h} > 1 \end{cases}$$

where $r$ is the distance from the center of a moving mesh cell to the center of a Cartesian cell and $h$ is the smoothing length:

$$h = (\text{Moving Mesh Cell Volume})^{1/3} \qquad (3)$$

Iterating over every cell in the moving mesh, this prescription ensures that all of the interpolated quantity is deposited into Cartesian cells. Afterward, we renormalize the entire Cartesian mesh to enforce the conservation of the interpolated quantity. Using this algorithm, the final density ($\rho$) of an interpolated quantity ($q$) in each Cartesian cell is given by:

$$\rho_i = \sum_{j=1}^{N} q_j W(|r_{ij}|, h_j) \qquad (4)$$

Here $i$ is the index of a Cartesian cell and $N$ is the number of moving mesh cells.

For our analysis, we used a Cartesian cell size that results in a snapshot resolution of $512^3$ cells for our interpolation. Appendix A shows comparisons between synthetic polarization maps with resolutions ranging from $256^3$ cells to $1024^3$. We chose to use the $512^3$ interpolation for our analysis, and show the convergence in Appendix A.1. It is import to note that it is only the interpolated snapshots that have these uniform, finer reso-lutions. The resolution of $512^3$ for this work was chosen as a compromise between resolution and computational requirements.

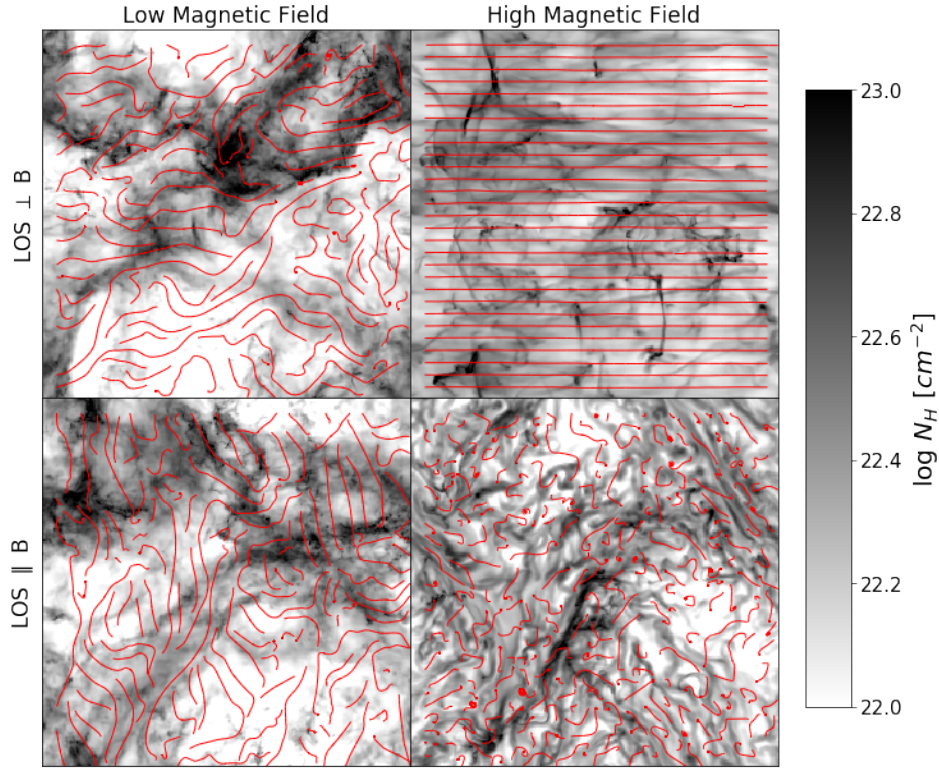

**Figure 1.** Snapshots of two different AREPO simulations from Mocz et al. (2017) from two different viewing orientations. The grey-scale maps in each plot show the column density ($N_H$) while the red lines are streamlines of the magnetic field in the plane-of-sky integrated along the observer's line-of-sight represented by each panel. The panels in the top row are viewed such that the mean magnetic field direction is parallel to the plane-of-sky while the bottom panels shows the mean magnetic field oriented toward the observer. The left column shows the simulation from run 1, which has the lowest initial average magnetic field strength (1.2 $\mu$G), while the right column shows the simulation from run 4, which has the highest initial magnetic field strength (120 $\mu$G).

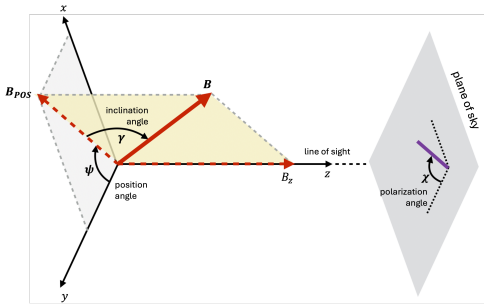

**Figure 2.** Adapted from Chen et al. (2019) The hypothetical observer sits in the plane on the right, facing the plane on the left. Their line-of-sight direction is denoted by the z coordinate. The mean magnetic field direction with inclination angle $\gamma$ with respect to the plane-of-sky, ranging from 0° ($B$ is parallel to the observer's plane-of-sky) to 90° ($B$ is parallel to the observers line-of-sight). The plane-of-sky magnetic field components, $B_x$ and $B_y$, come together to form the component of the magnetic field projected onto the plane-of-sky, $B_{POS}$, while the line-of-sight component is $B_z$.

### 3. SYNTHETIC DUST POLARIZATION MAPS

Linearly polarized thermal emission from dust grains is a common tracer of magnetic fields. Dust is generally well mixed with gas in the ISM and so polarized dust emission can be used to map magnetic fields in many different gas phases. Polarization can both provide information about the POS orientation of the magnetic field averaged along the dust column probed by the telescope beam and give insight into the inclination angle of the magnetic field if the maximum polarization fraction is known (Hildebrand 1988; Chen et al. 2019). For our study, we create synthetic polarization measurements and compare them at various viewing angles between simulation runs with different average magnetic field strengths (see Table 1). We generate our synthetic polarization observations by first calculating the Stokes parameters, $Q$ and $U$:

$$Q = \int n \frac{B_y^2 - B_x^2}{B^2} dz$$

$$U = \int n \frac{2 B_x B_y}{B^2} dz \quad (5)$$

as was done in Chen et al. (2019). If the observer's line-of-sight is along the z-axis, then the magnetic field components in the plane-of-sky are $B_x$ and $B_y$ and the magnitude of the magnetic field is given as $B$ as illustrated in Figure 2. Following these definitions, $Q$ and $U$ have the same units as column density, $\text{cm}^{-2}$, rather than intensity (MJy sr$^{-1}$). The angle of the polarization is given by

$$\chi = \frac{1}{2} \arctan\left(\frac{U}{Q}\right) \quad (6)$$

The local dispersion in $\chi$ is given by,

$$S^2(x,\delta) = \frac{\sum \Delta\chi^2(x,x_i)}{n_x} \quad (7)$$

where $x$ is the individual pixel and $n_x$ is the number of pixels $x_i$ that are within a distance of $\delta$ from $x$. For our analysis $\delta = 2$ pixels which corresponds to $0.02$ pc. We investigate this choice of $\delta$ in Appendix A.2.

We also calculate the polarization fraction as

$$p = p_0 \frac{\sqrt{Q^2 + U^2}}{N - p_0 N_2} \quad (8)$$

where $N = \int n dz$ is the column density integrated along the line-of-sight and $p_0$ is the internal polarization coefficient and which in observations can be estimated from the maximum polarization of the cloud (Chen et al. 2019). $N_2$ is a correction term used to consider reduced emission from inclined dust grains with smaller cross-sections and is calculated as

$$N_2 = \int n(\cos^2\gamma - \frac{2}{3}) dz \quad (9)$$

where $\gamma$ is the inclination of mean magnetic field direction. As in Chen et al. (2019), we take $p_0$ to be 0.1. Note that we are assuming that there is a uniform level of grain alignment throughout the cloud and that all variations in polarization are due to variations in the orientation of the magnetic field and not from changes in dust grain properties within the cloud. Under these assumptions, if there is little depolarization due to cancellation of Stokes Q and U from a disordered POS magnetic field, the inclination angle $\gamma$ of the magnetic field with respect to the POS (see Figure 2) becomes the dominant source of variation in polarization. This is because dust grains tend to rotate about the axis that gives the maximum moment of inertia, which is usually the short axis of the grain. Since the short axis of the grain preferentially aligns with the local magnetic field direction, the observer should see a maximum grain elongation if the magnetic field is parallel to the POS and no polarization if the magnetic field is along the line-of-sight. In this case, the inclination angle of the magnetic field can be estimated by measuring the polarization fraction given in Equation 8 and comparing it to the theoretical model:

$$p = \frac{p_0 \cos^2 \gamma}{1 + \frac{p_0}{3} - \frac{p_0 \cos^2 \gamma}{2}} \quad (10)$$

In this paper, we take $\gamma$ to be $0°$ when the magnetic field direction is parallel to the POS and $90°$ when it is parallel the observer's LOS as illustrated in Figure 2. If there is a signficantly disordered magnetic field component, the change in $p$ with $\gamma$ will be greatly reduced due to differences in inclination angle throughout the sight line and cancellation of the Stokes $Q$ and $U$ parameters due to field tangling.

### 3.1. *Generating Inclined Maps*

Each AREPO simulation run in Table 1 was generated such that the average magnetic field was parallel to the x-axis shown in the right panels of Figure 1. At each inclination, maps of Stokes Q and U and the resulting polarization fraction are calculated using Equations 5 and 8. In order to probe how the polarization observables change with mean inclination angle of the magnetic field $\gamma$, we create synthetic observations from different viewing angles. By convention when the magnetic field is entirely in the plane-of-sky and $\gamma = 0°$ the mean magnetic field direction is parallel to the x-axis. As $\gamma$ is increased, the simulation is rotated about the y-axis. Increasing $\gamma$ rotates the mean magnetic field until it reaches $\gamma = 90°$ where the magnetic field is parallel to the observer's line-of-sight. We analyzed each run at seven different inclination angles; $\gamma = 0°, 15°, 30°, 45°, 60°, 75°,$ and $90°$. The map for each of our quantities can be located in Appendix B.

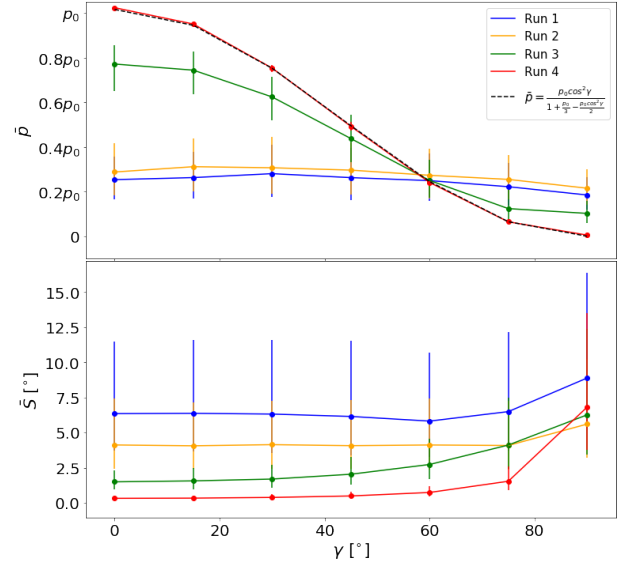

**Figure 3.** Top panel, median polarization fraction from the synthetic polarization map ($\bar{p}$) as a function of inclination angle ($\gamma$) of the magnetic field. Each $\bar{p}$ is given relative to $p_0$ which is the internal polarization coefficient in Equation 8. The vertical lines at each point show the interquartile range of $p$ values in each map. The predicted $p$ vs. $\gamma$ relation, Equation 10, is shown with the black dashed line in the top panel. The bottom panel shows the median of the polarization angle dispersion map $\tilde{S}$ and is plotted against $\gamma$. The vertical lines indicate the interquartile range.

To perform the analysis at various inclinations, the vector quantities, $\vec{B}$ and $\vec{v}$, were interpolated using SCIPY.SPATIAL.TRANSFORM.ROTATION about the y-axis given as $[0, 1, 0]$. This package employs spherical linear interpolation around a fixed axis, which in this case is the y-axis. The density, $\rho$, is rotated with SCIPY.NDIMAGE.INTERPOLATION.ROTATE in the x-z plane. This package employs spline interpolation of order 3 in a given plane.

### 3.2. *Dust Polarization Observables vs. Inclination Angle*

In the top panel of Figure 3 we show the median polarization fraction of our maps for the four different simulations as a function of $\gamma$. These are shown in Figure 3 as well as the theoretical result from Equation 10. As shown in Figure 3, the super-Alvénic cases (runs 1 and 2) show little variation in the polarization fraction, $p$, with inclination angle, $\gamma$. This suggests the mean magnetic field's orientation has little influence on $\bar{p}$, which is expected for cases where the magnetic energy density is weak compared to the kinetic energy of turbulent gas motions. When we plot $\bar{p}$ vs $\gamma$ for stronger magnetic field cases (runs 3 and 4) we see $\bar{p}$ follow the predicted $\cos^2 \gamma$ dependence from Equation 10, which is indicated with a dashed line in Figure 3. While the trans-Alvénic case (run 3) does exhibit cosine squared-like behavior as a function of $\gamma$, the measured polarization does not match the predicted maximum or minimum $p$ from Equation 10. The observed maximum $p$ is much lower than that of $p_0$ and the minimum higher than 0. In contrast, the sub-Alfvénic case (run 4) matches the predictions nearly identically. This is expected: since the magnetic field is strong, and the dispersion in polarization angle is minimal, there will be almost no decrease in $p$ due to magnetic field tangling within the beam, and the local magnetic field direction rarely deviates significantly from the mean magnetic field direction (see the upper right panel of Figure 1). Therefore, in the strong magnetic field case almost all variance in $p$ comes from $\gamma$.

One issue with solely using the polarization fraction to estimate 3-D magnetic field morphology is that it is dependent on the dust properties of the cloud. In particular, knowledge of the internal polarization coefficient $p_0$ from Equations 8 and 10 is needed. The maximum polarization fraction can depend on dust properties, cloud depth, the average inclination angle of the magnetic field, and may vary from cloud to cloud or within clouds. In our synthetic observations we assume a constant $p_0 = 0.1$ to match the assumptions in Chen et al. (2019), but the true $p_0$ is likely different within real clouds. It would be more useful for observers to estimate the inclination angle using a polarization observable that does not depend on dust grain properties and therefore offers a cleaner comparison of observations and simulations, such as the dispersion in the polarization angle, $S$. In the bottom panel of Figure 3 we show the median polarization angle dispersion, $S$ as a function of $\gamma$. We note that there is very little change of $S$ with $\gamma$, especially in the low inclinations ($\gamma \lesssim 60°$), so $S$ would seem to be less suitable for estimating the inclination of the magnetic field. Our observations of $\bar{S}$ do break the degeneracy between the very super-Alfvénic run 1 and the mildly super-Alfvénic run 2, which had very similar values of $\bar{p}$ in the top panel of Figure 3.

In Figure 4 we compare $p$ vs $S$ for different values of $\gamma$, and find that for the super-Alfvénic simulations (runs 1 and 2) there is little change in either measurement as a function of $\gamma$, leading to clustering in the $p$ vs $S$ parameter space. In the stronger field cases (runs 3 and 4, shown in the lower two panels of Figure 4) there is a definite relationship between $S$, $\gamma$, and $p$. In these synthetic observations, lower $\gamma$ (where the mean magnetic field is parallel to the POS) leads to higher $\bar{p}$ values, and smaller $\bar{S}$ values. In contrast, higher $\gamma$ (where the mean magnetic field is oriented close to the LOS) leads to lower $\bar{p}$ and higher $\bar{S}$. Note that for all simulations when viewing the cloud directly along the LOS we find extremely low $p$ and high $S$.

Using the polarization angle dispersion $S$ as a polarization measurable allows us to more easily compare simulations to actual polarization observations (without needing to know $p_0$). However, using $S$ to estimate inclina-tion still leaves us with two problems. First, in low-field strength simulations (runs 1 and 2) both $p$ and $S$ show very little dependence on inclination angle. Second, we are still unable to distinguish between cases of highly inclined, large $\gamma$, magnetic fields or magnetic fields that are just too weak to polarize thermal dust emission significantly. To break the second degeneracy we will suggest in the next section that it is necessary to estimate the average $\mathcal{M}_A$ of the gas.

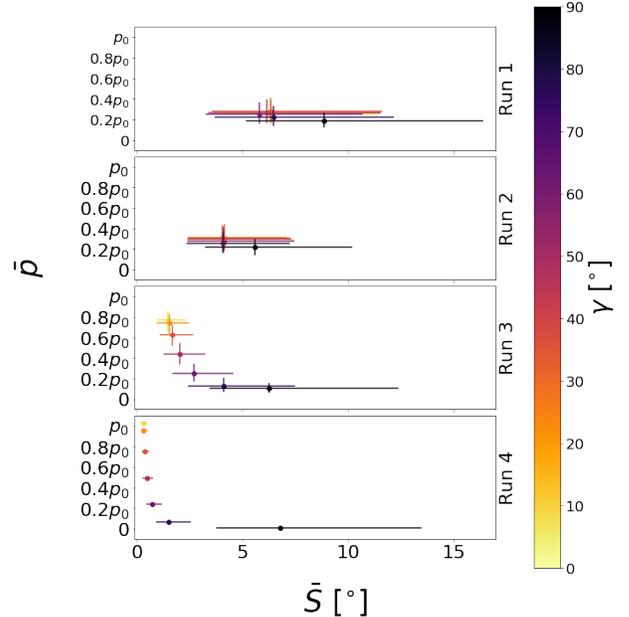

**Figure 4.** Distribution of $p$ and $S$ at different inclination angles $\gamma$, for the four simulations studied in this work. The closed circles indicate the median values of $p$ and $S$, while the vertical and horizontal lines show the interquartile distribution for $p$ and $S$, respectively. $\gamma$ is indicated by the color scale starting at $\gamma = 0°$ corresponding to the mean magnetic field being parallel to the POS, while $\gamma = 90°$ corresponds to the mean magnetic field oriented along the LOS. The average magnetic field strength increases from going from the top (run 1) to the bottom (run 4) panels.

## 4. THE ZEEMAN EFFECT

In the presence of a magnetic field, a spectral line will split into multiple components with an energy separation that is proportional to the magnetic field strength. Practically, it is only possible to measure the LOS component of the magnetic field strength through Zeeman splitting observations (Crutcher 2012; Crutcher & Kemball 2019) because magnetic fields in the interstellar medium are weak.

In order to calculate $\mathcal{M}_A$ from Equations 1 and 2 we would need the amplitude of the three-dimensional velocity field and magnetic field as well as a measurement of the gas density, which we cannot measure with Zeeman splitting observations. Instead, we define the LOS Alfvén Mach number

$$\mathcal{M}_{A,z} \approx \frac{\sqrt{3}\Delta v_z}{v_{A,z}}, \quad (11)$$

which is similar to $\mathcal{M}_A$ as defined in Equation 1, but calculated just from the LOS velocity dispersion $\Delta v_z$ and $v_{A,z}$, the Alfvén velocity calculated from the LOS com-ponent of the magnetic field.

$$v_{A,z} = \frac{|B_{LOS}|}{\sqrt{4\pi\bar{\rho}_z}}, \quad (12)$$

where $B_{LOS}$ is the density weighted LOS component of the magnetic field and $\bar{\rho}_z$ is the mean density along the observer's line-of-sight. All of these quantities can be estimated via Zeeman splitting observations. Note that the factor of $\sqrt{3}$ in Equation 11 comes from assuming that the 3-D velocity dispersion is isotropic. Under this assumption if the magnetic field is only along the line-of-sight then $\mathcal{M}_{A,z} = \mathcal{M}_A$. Note that $\mathcal{M}_{A,z}$ is not the true 3-D Alfvén Mach Number but what can be inferred through Zeeman splitting observations alone. It is possible for $\mathcal{M}_A$ to be less than $\mathcal{M}_{A,z}$ if the magnetic field is mostly orientated in the POS, or higher if the velocity dispersion is not isotropic and is larger in the POS than along the LOS.

We also note that Equation 12 requires a measurement of the gas density, $\rho$, which must be estimated from the line tracer used for the Zeeman splitting observations (i.e., HI 21-cm, OH, or CN lines). This may require assumptions about the gas pressure or the LOS depth of the cloud.

### 4.1. Inferred Alfven Mach Number from Synthetic Zeeman Observations.

To calculate $\mathcal{M}_{A,z}$ for our synthetic Zeeman observations we first define the density-weighted average line-of-sight velocity for each sightline:

$$\bar{v}_z = \frac{\int \rho v_z dz}{\int \rho dz} \quad (13)$$

where $v_z$ is the line-of-sight component of the gas velocity. Then we measure $\Delta v_z$ by calculating the standard deviation of the density weighted average of the velocity along the observer's LOS

$$\Delta v_z = \sqrt{\frac{\int \rho (\bar{v}_z - v_z)^2 dz}{\int \rho dz}}. \quad (14)$$

We then calculate $v_{A,z}$ for each pixel in the map using Equation 12, where $B_{LOS}$ is taken to be the density weighted average of the LOS component of the magnetic field for each voxel along the sight-line and $\bar{\rho}$ is taken to be the average density along the sightline. Our synthetic Zeeman observations then yield $\mathcal{M}_{A,z}$ by employing Equation 11. Maps of $\mathcal{M}_{A,z}$ can be found in Appendix B.

The left panel of Figure 5 shows estimated $\mathcal{M}_{A,z}$ vs $\gamma$ for all four simulations. The vertical lines at the location of each data point indicate the interquartile range of the $\mathcal{M}_{A,z}$ map while the points show the median $\mathcal{M}_{A,z}$ of the map. The blue horizontal lines present in both panels are the average 3-D $\mathcal{M}_A$ over the entire simulation snapshot volume.

From these synthetic observations, we see that a measurement of $\bar{\mathcal{M}}_{A,z} < 1$ always indicates that the cloud is on average sub-Alfvénic. If an observer measures that $\bar{\mathcal{M}}_{A,z} \gg 1$ then the cloud could be on average super-Alfvénc, trans-Alfvénic, or even sub-Alfvénic if it happens that $\gamma \approx 0°$.

In the right panel of Figure 5 we compare both $\mathcal{M}_{A,z}$ and the polarization angle dispersion $S$. We see that the four different simulation snapshots occupy different parts of the $\mathcal{M}_{A,z} - S$ parameter space. This suggests that a study of both Zeeman splitting measurements and dust polarization could be used to break the degeneracy in estimating $\mathcal{M}_A$ from $\mathcal{M}_{A,z}$ due to the inclination of the magnetic field, as will be discussed in the next section.

## 5. COMBINING DUST AND ZEEMAN INFORMATION

So far in this paper we have presented synthetic dust polarization observations and synthetic Zeeman splitting observations. In this section we will show how an observer could use both Zeeman and dust polarization observations to learn more about the magnetization of their target cloud.

### 5.1. The Easy PZ Method for Estimating $\mathcal{M}_A$

In the right panel of Figure 5 we compare $\mathcal{M}_{A,z}$ with $S$ and note that different simulations occupy different regions of the parameter space of the observables $\mathcal{M}_{A,z}$ and $S$. We note that for stronger magnetic fields (runs 3 and 4), $\mathcal{M}_{A,z}$ shows dependence on $\gamma$ while there is little dependence on $\gamma$ in the weakest cases (runs 1 and 2). For all $\gamma$, the $S$ vs $\mathcal{M}_{A,z}$ observations are clustered within a narrow range of the parameter space for both runs 1 and 2. For the trans-Alfvénic simulation, run 3, the estimated $\mathcal{M}_{A,z}$ and $S$ both have a strong dependence on $\gamma$. The $\mathcal{M}_{A,z}$ decreases with increasing $\gamma$ while $S$ increases. In the sub-Alfvénic run 4, we either observe $M_{A,z} < 1$ when $\gamma \geq 30°$, or for higher inclination angles ($\gamma \leq 60°$) we observe a very low median polarization angle dispersion ($S \leq 1°$). There is not much variation in $\mathcal{M}_{A,z}$ with $\gamma$ as we observe $\bar{\mathcal{M}}_{A,z} \leq 1$ for $\gamma \geq 15°$. Consequently, should a series of Zeeman and dust polarization observations across a cloud lead to a cluster of observations where $\mathcal{M}_{A,z} \gg 1$ and $S \gg 1°$ then the observer could infer the gas is super-Alfvénic. Although one would not be able to infer much about the inclination angle in this case, these observations would imply that the 3-D magnetic field is quite disordered and is energetically sub-dominant to the kinetic energy of the turbulent gas motions. However, if the observations show $\mathcal{M}_{A,z} \leq 1$ or $\mathcal{M}_{A,z} > 1$ and $S \lessapprox 2°$, then the observer could infer that the cloud is sub-Alfvénic or trans-Alfvénic.

Using these synthetic observations, we propose a method to infer whether the cloud is super-Alfvénic, trans-Alfvénic, or sub-Alfvénic as shown schematically in Figure 6. First, our hypothetical observer would take many Zeeman observations across a cloud and estimate $\bar{\mathcal{M}}_{A,z}$. Depending on the value of $\bar{\mathcal{M}}_{A,z}$, there are different cases which we outline in the following subsections.

#### 5.1.1. $\mathcal{M}_{A,z} \ll 1$

If $\mathcal{M}_{A,z} \ll 1$ (right path of Figure 6), and the cloud is measured to be sub-Alfvénic, then the observer can conclude that the cloud is indeed sub-Alfvénic. In this set of four simulation snaphots viewed from seven different inclination angles $\gamma$ there was no case where we measured $\bar{\mathcal{M}}_{A,z} < 1$ and the 3-D $\bar{\mathcal{M}}_A$ was actually trans- or super-Alfvénic. If the cloud is sub-Alfvénic, then $p$ should depend strongly on $\gamma$. If it is possible to estimate $p_0$, then $\gamma$ could be estimated using methods similar to those outlined in Chen et al. (2019).

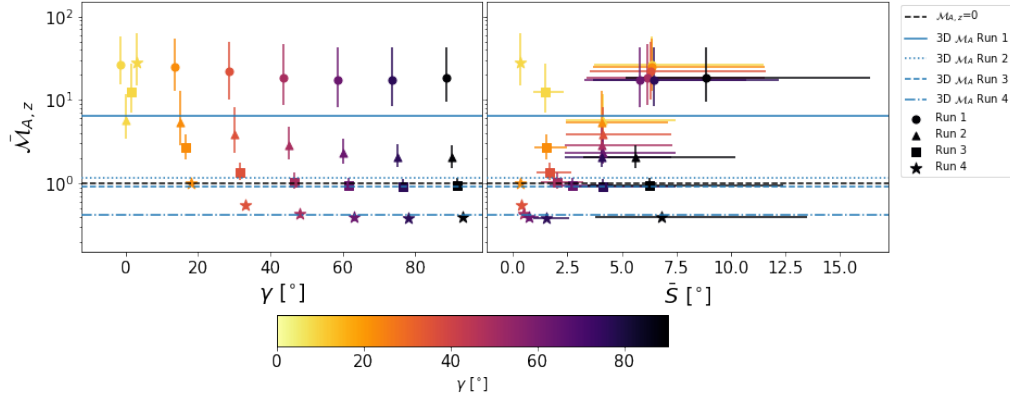

**Figure 5.** Left panel $\bar{\mathcal{M}}_{A,z}$ as a function of $\gamma$. The vertical spread indicates the interquartile range of $\mathcal{M}_{A,z}$. So that the interquartile ranges do not overlap, run 1 has had its x-position offset by $-1.5°$, run 3 by $1.5°$, and run 4 by $3°$. **Right panel** $\bar{\mathcal{M}}_{A,z}$ is plotted as a function of $\bar{S}$. The horizontal blue lines in both panels represent the average 3-D Mach Number of each snap shot. The black dotted line indicates $\mathcal{M}_{A,z} = 1$.

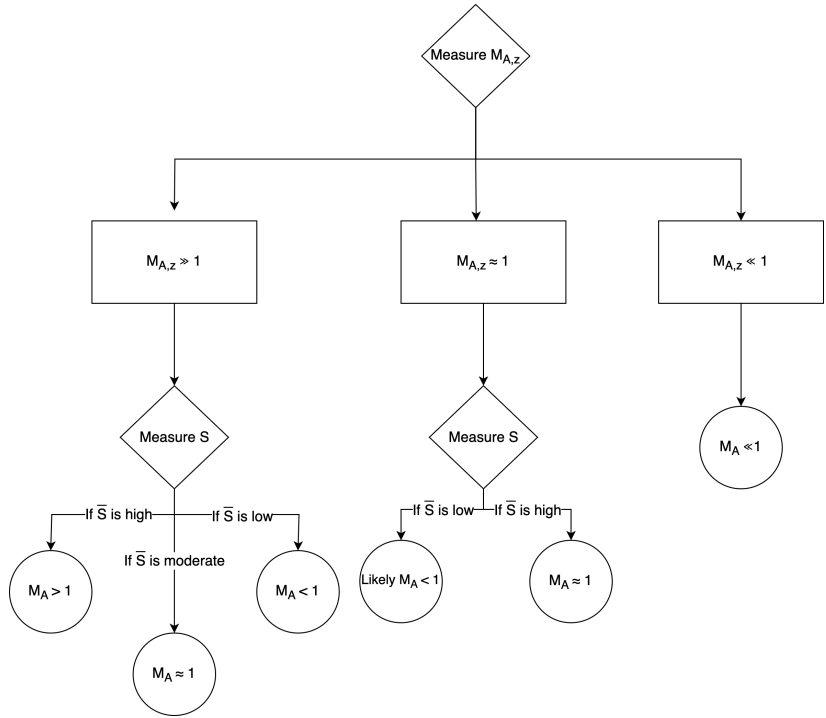

**Figure 6.** Flowchart outlining our suggested processes for comparing the Zeeman and dust polarization observations in order to infer whether a cloud is on average sub-, trans-, or super-Alfvénic. The diamonds represent making a measurement, the rectangles represent drawing an inference from the measurement, and the circles represent a conclusion to draw from one's path through the tree. An observer would first take a Zeeman measurement then determine the median value for the $\mathcal{M}_{A,z}$. If $\mathcal{M}_{A,z}$ is $< 1$ then one can conclude that $\mathcal{M}_A$ is also $< 1$. Otherwise, the observer will have to make and compare their $S$ measurements and follow the appropriate paths based on those conclusions.

### 5.1.2. $\mathcal{M}_{A,z} \approx 1$

If $\mathcal{M}_{A,z} \approx 1$ the cloud could be sub-Alfvénic or trans-Alfvénic (center path of Figure 6). If $S$ is high, $S \gtrsim 3°$ in these simulations, then the cloud is likely to be trans-Alfvénic. If $S$ is low, $S \lesssim 1°$ in these simulations, then the cloud is likely to be sub-Alfvénic. In this case, $p$ will be strongly dependent on $\gamma$ and it may be possible to estimate the inclination angle of the magnetic field, and therefore the 3-D $\mathcal{M}_A$.

### 5.1.3. $\mathcal{M}_{A,z} \gg 1$

If $\mathcal{M}_{A,z} \gg 1$, we again need dust polarization observables to constrain the three dimensional Alfvén proper-ties (left path of Figure 6). From our synthetic observations we find that if $S$ is low ($\bar{S} \lesssim 1°$) that implies that the magnetic field is mostly parallel to the plane-of-sky (low $\gamma$) and the cloud is actually sub-Alfvénic. If $S$ is high ($\bar{S} \gtrsim 3°$), we find that the magnetic field is highly disordered in the POS which suggests that the cloud is super-Alfvénic. For the super-Alfvénic simulations the magnetic field is so disordered that in most cases there is almost no dependence of $\mathcal{M}_{A,z}$ or $S$ on inclination angle, likely because the mean field is quite weak compared to the turbulent component. If $S$ is moderate (in our simulations if $\bar{S} \geq 1°$ and $\leq 3°$), the cloud is likely trans-Alfvénic. As discussed in the previous sub-sections it may be possible to estimate $\gamma$ for a trans-Alfvénic or sub-Alfvénic cloud in these conditions.

## 6. DISCUSSION

In this paper, we used numerical simulations to outline a methodology for comparing Zeeman and dust polarization information in order to gain insights into the inclination angle and field strength. We discuss some of the complications of applying these methods to real observational data below.

### 6.1. *Application to Observational Data*

First, we note that our synthetic Zeeman measurements assume a "perfect" spectral line that traces the entire gas column. To mimic a Zeeman measurement, our $B_{LOS}$ is the density-weighted mean of the LOS magnetic field component. To better interpret our synthetic tracer, we have plotted $\bar{B}_{LOS}$ as a function of $\gamma$ in the left panel of Figure 7 and as a function of $\bar{S}$ in the right panel. We note that in the sub-Alfvénic run 4, and to a lesser extent the trans-Alfvénic run 3, $B_{LOS}$ begins to match the 3-D $B$ at $\gamma \approx 75°$. However in the two super-Alfvénic cases exhibit little to no relationship with $\gamma$. In runs 1 and 2 $\frac{\bar{B}_{LOS}}{\bar{B}_{3D}}$ is approximately 0.5 for all values of $\gamma$. This is consistent with observational approximations often used, where for a set of randomly ordered magnetic field orientations the mean 3-D field strength is assumed to be twice the mean $B_{LOS}$ (Crutcher 2012).

The simple prescription in 5.1 might not be immediately applicable to real polarization observations, as we have ignored dust grain physics such as grain alignment and temperature variations which could cause dust polarization observations to sample some dust populations along a sight-line more than others. Additionally, we have not generated synthetic Zeeman observations for different spectral lines. We have instead assumed an idealized Zeeman measurement that traces the magnetic field in the same gas as probed by our dust polarization observations. There are also some physics that are missing from the simulations that we have used. In real molecular clouds, there is probably not just one mean field direction, but a more complicated large-scale magnetic field structure. More realistic cloud simulations would also include feedback from massive stars, as well as gas heating and cooling. These simulations also assume flux frozen, idealized MHD conditions.

We are presenting this hypothetical method for estimating cloud magnetization by comparing Zeeman and dust polarization as an initial simulation exploration. This analysis method should be tested on more realistic synthetic observations. We also note that, for most observations, Zeeman measurements are detected for a few individual sight-lines within a cloud rather than measuring $B_{LOS}$ over an entire cloud as we have assumed in our analysis.

Observers are also unable to rotate their LOS around molecular clouds the same way theorists can when making synthetic observations of simulations. However, one can still trace three-dimensional magnetic field structure of a molecular cloud by employing our method to a series of sub-regions of the cloud. The magnetic field direction of real molecular clouds often show large-scale changes in both POS direction and inclination angle (Tahani et al. 2019), in contrast to the single mean inclination angle of our simulations. The changing direction of the mean magnetic field could allow an observer to probe many inclination angles of the magnetic field within one gas cloud.

Future work will also include comparing Zeeman and dust polarization measurements to each other, where available, in real clouds. By comparing to works such as the catalogue of Zeeman splitting observations from Crutcher (2012), linear dust polarization observations from telescopes such as BLASTPol, (Fissel et al. 2016), *Planck* (Planck Collaboration et al. 2015a), and SOFIA/HAWC+ (Harper et al. 2018b) we could attempt to probe the 3-D orientation and strength of well-studied molecular clouds. The cases explored in the synthetic dust polarization study by Chen et al. (2019) are broadly trans-Alfvénic. However, some molecular clouds appear to be super-Alfvénic (Federrath et al. 2016). Generating synthetic polarization observation by using a post-processing code that inclcudes radiative transfer, such as POLARIS (Reissl et al. 2016), on these simulations would produce more realistic Zeeman observations that can be directly compared with observations.

## 7. CONCLUSIONS

We compared two well-studied magnetic field tracers (dust polarization and the Zeeman effect) using synthetic observations of simulated star-forming molecular clouds. Our clouds are a suite of four isothermal AREPO simulations from Mocz et al. (2017). Each simulation has an initial sonic Mach number of $\mathcal{M}_s = 10$ while the initial magnetic field strength, and therefore Alfvén Mach Numbers, is different in each simulation. We generated our synthetic dust polarization observations assuming uniform dust properties and grain alignment efficiencies. We created synthetic polarization observations from seven different viewing angles, $\gamma$, with respect to the mean magnetic field, such that for $\gamma = 0°$ the mean magnetic field is parallel to the plane of sky, and for $\gamma = 90°$ the mean magnetic field is along the line of sight. From our synthetic dust polarization maps we calculated the polarization fraction $p$ of the emission and the local angular dispersion of the magnetic field $S$. Additionally, we have made idealized synthetic Zeeman splitting observations by calculating a density-weighted $B_{LOS}$ for each sightline. In order to probe the energetic importance of the magnetic field, we have defined an $\mathcal{M}_{A,z}$ which is the Alfvén Mach Number calculated from just the line-of-sight quantities that can be measured with Zeeman observations.

Our main conclusions are listed as follows:

1. We confirm that dust polarization fraction measurements are an effective probe of the inclination angle of the magnetic field (Equation 10) if the cloud is on average trans- or sub-Alfvénic. $p$ in trans-Alfvénic and sub-Alfvénic clouds has a strong dependance on $\gamma$. If $\mathcal{M}_{A,z}$ is on average $\gg 1$, then it is difficult to determine $\gamma$ but in those cases the mean field is likely weak compared to the disordered, turbulent component of the magnetic field.

2. Use of $p$ alone makes it difficult for an observation to distinguish between a super-Alfvénic cloud (with

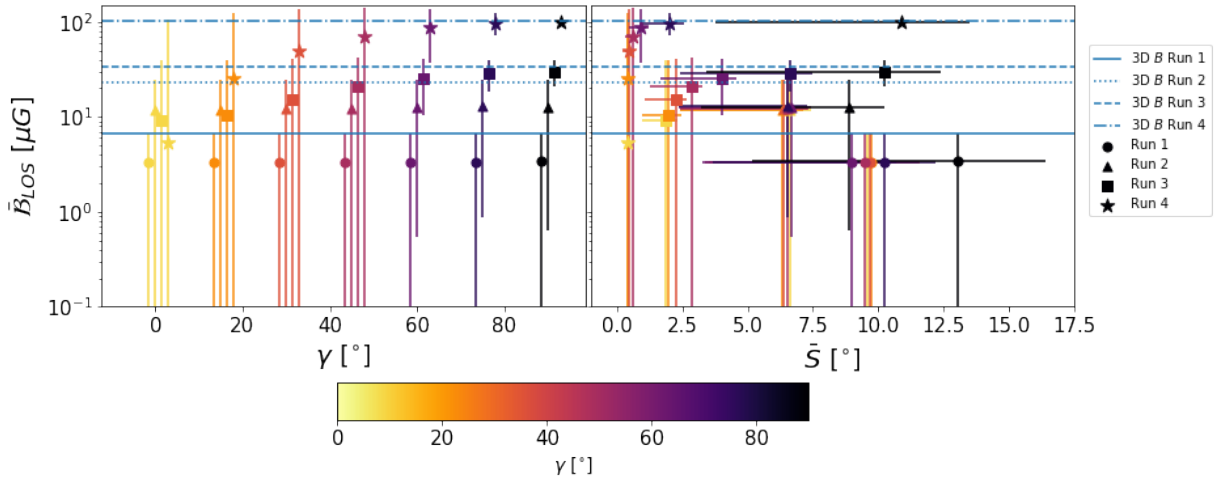

**Figure 7.** **Left panel:** $\bar{B}_{LOS}$ is calculated as a function of $\gamma$. Each point is the median for the map at that inclination. The vertical spread is given as the interquartile range of $B_{LOS}$. To prevent the interquartile ranges from overlapping, run 1 has been given a horizontal offset of $-1.5°$, run 3 an offset of $1.5°$, and run 4 an offset of $3°$. **Right panel:** $\bar{B}_{LOS}$ is plotted as a function of $\bar{S}$. The blue horizontal lines in each panel are the density weighted averaged 3-D $B$-field strength.

any $\gamma$) or a highly inclined, trans-Alfvénic or sub-Alfvénic cloud (where $\gamma \gtrsim 60°$). Additionally, by examining $p$ distributions alone we are unable to distinguish between the mildy super-Alfvénic and very super-Alfvénic runs.

3. Sub-Alfvénic and super-Alfvénic clouds will occupy different regions of the $\mathcal{M}_{A,z}$ - $S$ parameter space as shown in Figure 5. By comparing the synthetic observations at different $\gamma$, we find that super-Alfvénic clouds will show little dependence on $\gamma$ in the $\mathcal{M}_{A,z}$ - $S$ parameter space. The super-Alfvénic clouds will cluster in the regions with high $S$ and high $\mathcal{M}_{A,z}$. In contrast, measurements of sub-Alfvénic clouds are much more dependent on the observer's line of sight.

4. In the trans-Alfvénic and sub-Alfvénic cases, there is a dependence of $\mathcal{M}_{A,z}$ and $S$ on $\gamma$. In general, as the magnetic field direction becomes more inclined with respect to the plane-of-sky, $S$ will increase while $\mathcal{M}_{A,z}$ decreases. The amplitude of this change (and the constraint of the spread in which these values take on) varies between the trans-Alfvénic and sub-Alfvénic cases. The sub-Alfvénic case has the most dependence on $\gamma$ with low inclinations leading to high $\mathcal{M}_{A,z}$ (suggesting the cloud is super-Alfvénic when it is not) and low $S$.

5. We find that in our simulation suite, the Alfvén Mach Number and, in some cases, the 3-D magnetic field structure can be estimated by employing the Easy PZ Method which is illustrated in Figure 6. First, an observer starts by taking Zeeman observations of their target and concluding an estimate for $\mathcal{M}_{A,z}$. If $\mathcal{M}_{A,z} < 1$, then the cloud is likely sub-Alfvénic on average. Otherwise, measurements of $S$ need to be taken. Then $S$ and $\mathcal{M}_{A,z}$ can be compared to each other to infer the Alfvén Mach Number. If it is likely that $\mathcal{M}_A \lessapprox 1$, then the dust polarization fraction $p$ could possibly be used to estimate the inclination angle of the magnetic field.

B.B. is grateful for generous support from the Flatiron Institute, the David and Lucile Packard Foundation, and the Alfred P. Sloan Foundation. The Flatiron Institute is a division of the Simons Foundation. L.M.F acknowledges support from the National Science and Engineering Research Council (NSERC) through Discovery Grant RGPIN/06266-2020, and funding through the Queen's University Research Initiation Grant. S.E.C. acknowledges support from NSF award AST-2106607, NASA award 80NSSC23K0972, and an Alfred P. Sloan Research Fellowship.

## APPENDIX

### A. RESOLUTION STUDIES

#### A.1. *Interpolated Run Resolution Convergence*

The pre-interpolation snapshots used in our analysis have an effective resolution of $256^3$ cells. We have performed a resolution study to see the effects of resolution on the synthetic polarization observations. As discussed in 2.1, our interpolation code can be used to generate maps of higher and lower resolution than $256^3$. We show a resolution study using the extreme magnetic field simulations, (runs 1 and 4) at varying resolutions from $256^3$ to $1024^3$ in Figure 8. Each line corresponds to the running mean of the column density of each snapshot at the various resolutions. The shading represents 1-$\sigma$ deviation from the mean. Promisingly, the shape of the running mean lines of the column density match well with each other for a given magnetic field strength and line-of-sight, indicating that resolution differences alone will not change the overall measurement of the polarization fraction. Each of the running means with the same magnetic field are well within 1-$\sigma$ of each other. Finally, as expected, the higher-resolution interpolations have a wider range of column densities.

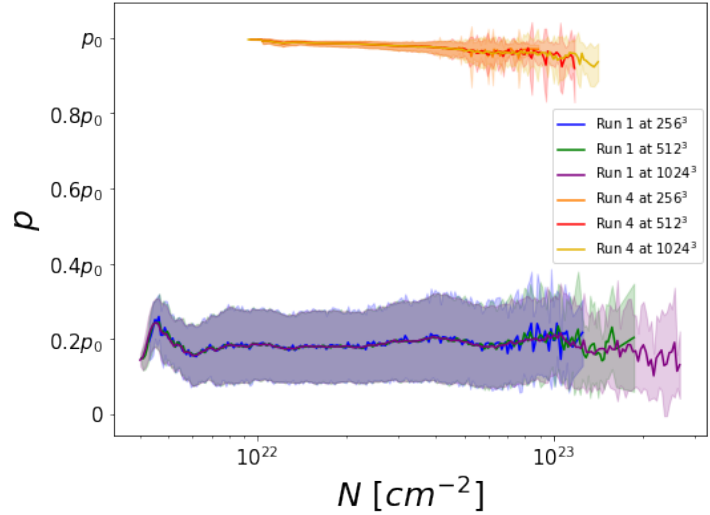

**Figure 8.** The original snapshots and the simulations had an effective resolution of $256^3$ as discussed in Section 2. To accurately apply particle information we interpolated a new grid onto our simulations to properly do our synthetic observations (see 2.1). We can, in principle, apply any resolution on top of the native grid however, each comes with its own kind of limitation. As can be seen in this figure, if we choose a resolution that is too small we lose a significant amount of the possible dynamic range in column density. However, the larger resolutions will require more computing power to analyze and perform our synthetic observations. To balance these issues, we performed our analysis at a resolution of $512^3$ which is a solid middle ground between dynamic range and computation requirements. All of these synthetic measurements were made at $\gamma = 0°$.

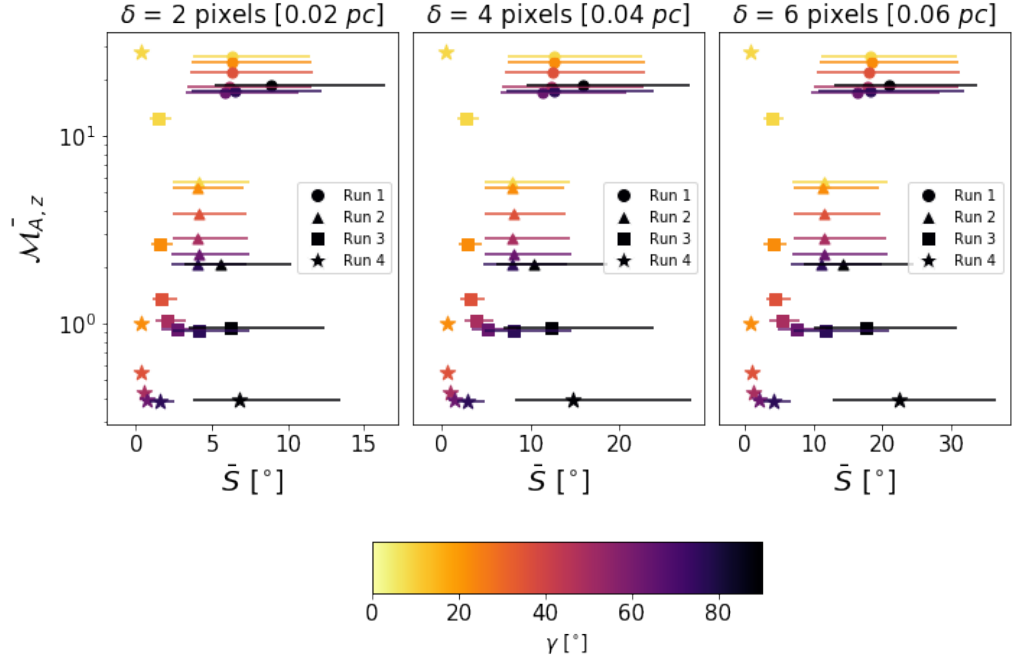

**Figure 9.** We recreate the polarization angle dispersion, $S$, observations with varying $\delta$ ranging from $\delta = 2$ pixels in the left panel to $\delta = 6$ pixels in the right panel. The position of the points are given by the median of the maps of $S$ and $\mathcal{M}_{A,z}$ to mimic Figure 5. The horizontal spread is given by the interquartile range of the $S$ maps.

### A.2. Study of $\delta$ Choice for $S$ Maps

The dispersion in polarization angle, $S$, at a given pixel is defined as the average difference between polarization vectors at each other pixel within a distance $\delta$ (Planck Collaboration et al. 2015b; Fissel et al. 2016). In Equation 7 we chose $\delta = 2$ pixels. We investigate increasing $\delta$ to 4 and 6 pixels corresponding to 0.04 pc and 0.06 pc respectively in Figure 9. We note that while the change in $\delta$ adjusts the absolute values that the $S$ maps take on (affecting the median and interquartile ranges of those maps), the trends between $S$, $\mathcal{M}_{A,z}$, $\gamma$, and $B$ strength outlined in Section 5 remain the same.

## B. MAPS OF SYNTHETIC OBSERVATIONS

For our analysis, we compared and rotated four different runs tabulated in Table 1. Further detail on the simulations themselves can be found in Mocz et al. (2017). The simulations are rotated at various inclination angles and the synthetic observations are taken. We have seven inclination angles ranging from $0°$ to $90°$ in intervals of $15°$. This gets us a total of 28 maps for each observable. For this analysis, our main three observables that we have tried to replicate are the polarization fraction from thermal dust emission, $p$, given by Equation 8, the dispersion in polarization angle, $S$, given by Equation 7, and the line-of-sight Alfvén Mach Number given by Equation 11. All 28 maps for $p$ are found Figure 10, those for $S$ can be found in Figure 11, and $\mathcal{M}_{A,z}$ in Figure 12.

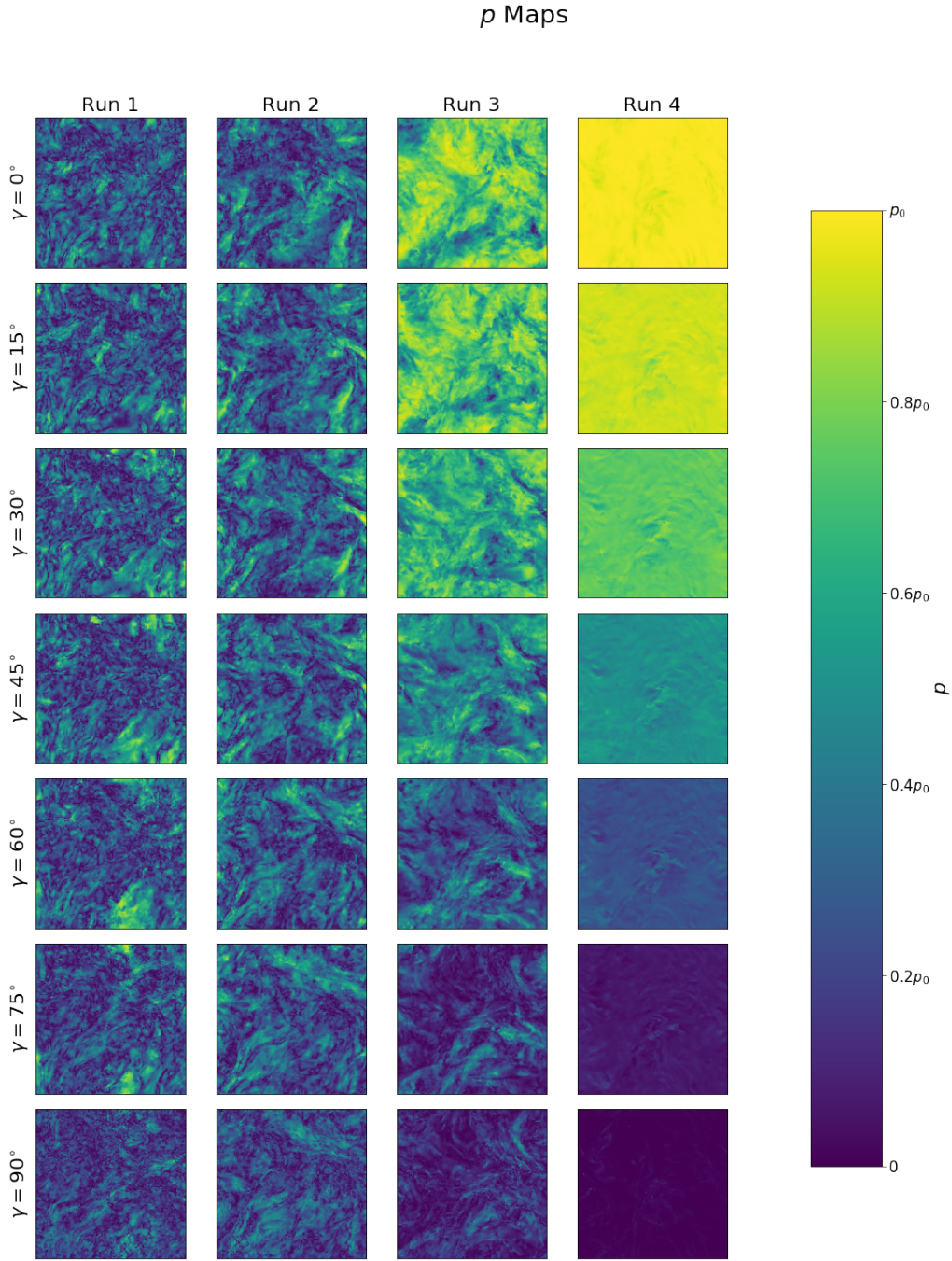

**Figure 10.** Each row represents one of the seven inclinations studied in this work. Each column represents one of the four runs described in Table 1. Each panel is a map of $p$ for a run at a given inclination.

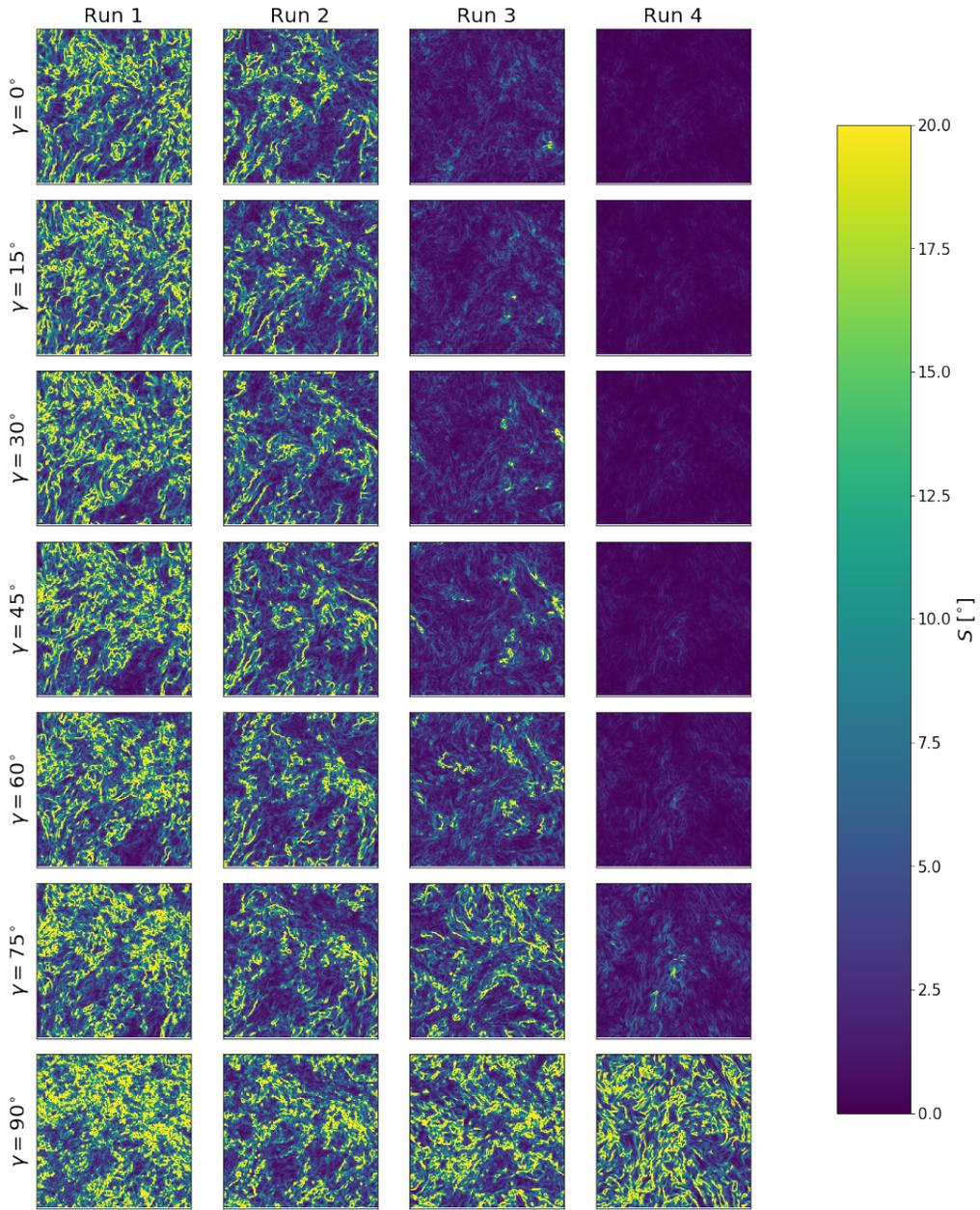

**Figure 11.** Each row represents one of the seven inclinations studied in this work. Each column represents one of the four runs described in Table 1. Each panel is a map of $S$ for a run at a given inclination.

## $\mathcal{M}_{A,z}$ Maps

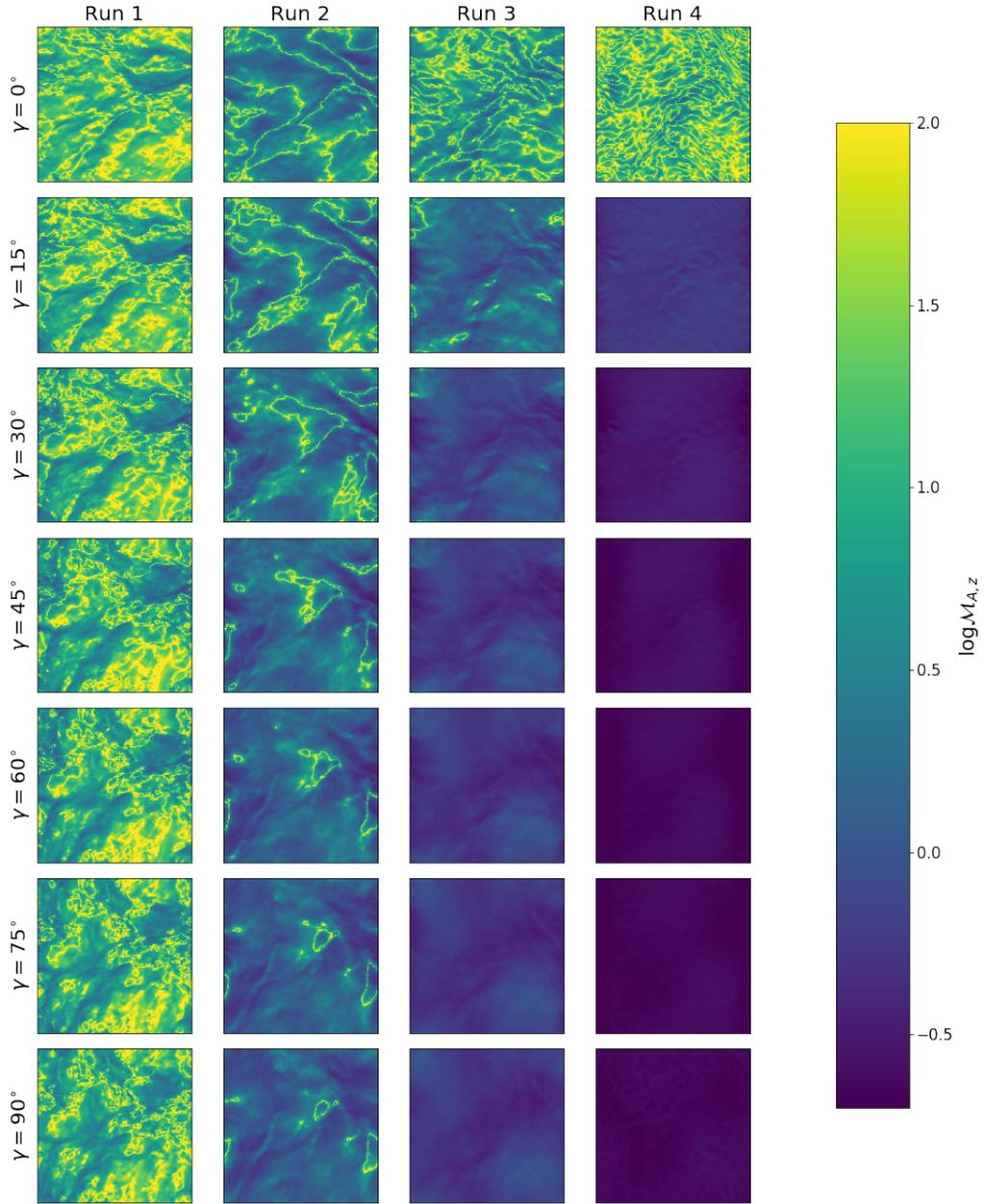

**Figure 12.** Each row represents one of the seven inclinations studied in this work. Each column represents one of the four runs described in Table 1. Each panel is a map of $\mathcal{M}_{A,z}$ for a run at a given inclination.